# SKA Aperture Array Verification System: Electromagnetic modeling and beam pattern measurements using a micro UAV

E. de Lera Acedo[1], P. Bolli[2], F. Paonessa[3], G. Virone[3], E. Colin-Beltran[4], N. Razavi-Ghods[1], I. Aicardi[5], A. Lingua[5], P. Maschio[5], J. Monari[6], G. Naldi[6], M. Piras[5], G. Pupillo[6]

[1]Cavendish Laboratory, Department of Physics, University of Cambridge, Cambridge CB3 0HE, UK
email: eloy@mrao.cam.ac.uk

[2]National Institute for Astrophysics, Arcetri Astrophysical Observatory, Florence, Italy

[3]National Research Council, Institute of Electronics, Information and Telecommunication, Turin, Italy

[4]Wireless Communications Research, Intel Corporation, Guadalajara, Mexico

[5]Polytechnic of Turin , Department of Environment, Land and Infrastructure Engineering, Turin, Italy

[6]National Institute for Astrophysics, Institute of Radio Astronomy, Bologna, Italy

In this paper we present the electromagnetic modeling and beam pattern measurements of a 16-elements ultra wideband sparse random test array for the low frequency instrument of the Square Kilometer Array telescope. We discuss the importance of a small array test platform for the development of technologies and techniques towards the final telescope, highlighting the most relevant aspects of its design. We also describe the electromagnetic simulations and modeling work as well as the embedded-element and array pattern measurements using an Unmanned Aerial Vehicle system. The latter are helpful both for the validation of the models and the design as well as for the future instrumental calibration of the telescope thanks to the stable, accurate and strong radio frequency signal transmitted by the UAV. At this stage of the design, these measurements have shown a general agreement between experimental results and numerical data and have revealed the localized effect of un-calibrated cable lengths in the inner side-lobes of the array pattern.

1. INTRODUCTION

The Square Kilometer Array (SKA) [1] is a radio interferometer envisaged to be the largest and most sensitive radio telescope in the world at meter and centimeter wavelengths by the time of its completion in the 2020s. This initial phase-I will include an aperture array instrument called SKA1-LOW (see Fig. 1), consisting of 512 phased array stations with 256 antenna elements in each station [2]. We would like to emphasize that [2] is an older (v6) but publically available version of the SKA1 L1 requirements. Version 10 is already in use by the project currently. Phase-II will follow by the mid/end of the 2020s with an order of magnitude more elements and sensitivity. Phase-I of the SKA is aiming to detect and image the Epoch of Re-ionization (EoR) and to study the Cosmic Dawn (CD) [3]. These epochs of the history of the Universe comprise the time from when darkness was prevailing in the Universe after the Big Bang until the end of cosmic re-ionization following the formation of the first stars and galaxies.

The SKA1-LOW, with it unprecedented sensitivity, will survey the sky of the southern hemisphere (it will be located in the deserts of Western Australia) faster than ever before. The SKA1-LOW will operate from 50 to 350 MHz (and possibly up to 500 MHz) covering both the red-shifted hydrogen emission from the CD and the EoR and accommodating other sciences such as gravitational waves and relativity through the study of all the pulsars in our Galaxy.

In order to meet its demanding requirements [2], the SKA1-LOW requires state-of-the-art technology in several fields, including high performance computing, digital processing, fiber optics communications, etc. The antenna arrays are not exempt from these demanding requirements. Ultra wideband (7:1 or more) antenna elements covering a 90 degrees field of view around zenith with extraordinary sensitivity are required. The SKA Log-periodic Antenna (SKALA), (described in section II and in [4] in more detail) has been designed for the purpose. Furthermore, it has been identified that the use of sparse random arrays brings numerous advantages [5],[6] for the astronomical application in question. In 2012, a prototype array, called Aperture Array Verification System 0 (AAVS0), was built at the Mullard Radio Astronomy Observatory (MRAO), south west of Cambridge (UK) to help the development of the different technologies being designed by the Aperture Array Design Construction (AADC) consortium for the SKA. This consortium is led by ASTRON (Netherlands) and has 6 member countries: Netherlands, UK, Italy, Australia, Malta and China. This array, after the measurement campaign described in this paper, has been recently upgraded with version 2 of the SKALA elements [7] and a digital back-end.

One of the greatest challenges for the aperture arrays of the SKA1-LOW is calibration, which in turn requires an accurate knowledge of the antenna/array beam pattern response as well as of the amplitude/phase response of the receiver chains. It has been discussed in recent papers [8-10] the importance of including accurate knowledge of antenna models in the calibration and imaging pipelines in order to achieve high fidelity images. In those papers simulated models are used. The measurements presented in this paper, although amplitude only and for a small array, are a step forward in the characterization and validation of the type of beam models that will need to be used in future radio instruments. Accurate knowledge of the embedded patterns will be used for the calibration and prediction of station beams and therefore will result in high dynamic range images [8]. Some of the reasons why calibration is harder than for the case of dishes include the changing side-lobe pattern with scan angle, the effects of mutual coupling, the electrical size of the array precluding the use of anechoic chambers and finally the low sensitivity of each single antenna of the array. The later imposes a challenge if one wants to characterize the receiver chains by using astronomical calibrators. The millions of sky sources contributing to this signal can be captured through the changing side-lobes and because of several effects (mutual coupling, temperature and environmental effects, effects of soil, etc.) will suffer changes in time and frequency. Therefore, both accurate predictions of the beam models through electromagnetic simulations and measurements of the beams are necessary. For instance, the spectral stability required by the SKA for the post-station calibration is 0.4% at 100 MHz (equivalent to few hundredths of dB) compared to the parameterized beam model. This is a very challenging figure, which can be reached only with a very deep understanding of all the single elements of the system, including therefore the antenna patterns.

This paper focuses on the use of an Unmanned Aerial Vehicle (UAV) test system to characterize onsite the AAVS0. Such a system has been developed, within the SKA framework, by a collaboration between the Polytechnic of Turin, the Italian National Council of Research and the Italian National Institute for Astrophysics [11]. It allows the measurement of the far-field radiation pattern of low-frequency antennas [12], [13] and small arrays in the real operative environment. In particular, it has already been applied to instrumentally verify (co-polar only) and calibrate a regular 3 x 3 Vivaldi array at 408 MHz [14],[15]. With respect to this previous campaign, the following primary objectives aimed the AAVS0 session: *i)* quantifying the antenna coupling effects by comparing the single-element pattern to the embedded ones; *ii)* measuring both co- and cross-polar components in the overall sky-coverage (±45° from zenith) which also allow to assess the polarization characteristics of the array with respect to scan angle; *iii)* validating the system at different frequencies down to 50 MHz (where the ground contribution increases significantly); *iv)* performing a raster scan to obtain 2D map of the pattern. Additionally, AAVS0 consists of a random sparse distribution of SKALA antennas, and this makes the campaign more significant in view of SKA1-LOW. Finally, with the measurements described in this paper we have being able to show the effect of un-calibrated phase in the RF chains in the inner side lobes of a random phased array while the main beam and far-out side lobes are almost unaffected, as predicted by theory.

In this paper we discuss the tests and validation of random ultra-wideband array, array patterns and embedded element patterns. These will be able to inform the SKA requirements [2] 2165 (polarization purity), 2622 (sensitivity off zenith angles), 2135-38 and 2814-15 (sensitivity per polarization) and 2629 (station beam stability). At the moment they are measurements of the amplitude of the patterns only, but they can be already used to estimate sensitivity (via the validation of the simulated models) as well as the other parameters mentioned.

Section 2 of this paper describes the antenna element under test, section 3 focuses on the array system design. Section 4 describes the electromagnetic modeling of AAVS0. Section 5 is then dedicated to the UAV pattern measurement system used in the array characterization. Section 6 discusses the results and in section 7 we draw some conclusions.

## 2. THE SKALA ELEMENT

SKALA is a linear dual-polarized 9-dipole log-periodic antenna (see Fig. 2) that has been designed to meet the demanding requirements of the SKA1-LOW telescope [3]. The design targeted maximum sensitivity, measured as the ratio of the effective aperture over the system noise (including the sky noise) across the frequency band and field of view specified for the SKA1-

LOW instrument while maintaining high polarization isolation [2]. For SKA, a metallic mesh grid will be placed under the antenna elements. The antenna is a high gain element matched to a Low Noise Amplifier (LNA). The LNA, designed by the University of Cambridge in collaboration with Cambridge Consultants Ltd, is a pseudo-differential LNA based on a dual AVAGO transistor chip in version 1 and a Qorvo transistor in version 2. The pseudo-differential first stage, designed as such for minimum receiver match noise [4] is followed by a wideband balun and a second stage amplifier. The LNA boards (1 per polarization) are located on an enclosure at the top of the antenna, right at the feed point. These boards will be followed, for SKA1-LOW, by a Radio Frequency over Fiber (RFoF) transmitter and a fiber output plus a twisted pair for power supply. Currently, our test array uses copper to feed power to the LNAs and to transmit the RF output signals to the back end receiver.

More than 70 copies of this antenna based on a wire bending technique (v1) have been built to date and have been tested in different laboratories across the world. Version 2 of SKALA has now been designed and built as a prototype. The only differences between SKALA-1 and -2 are on the mechanical design and LNA design, which in v2 have been improved for mass production. The EM response of the antenna has been shown to be almost identical between both versions [7]. Some preliminary tests on the SKALA single element performed with the UAV system have already been reported in [12].

3. THE AAVS0 TEST SYSTEM

The AAVS0 array (see Fig. 3) was conceived as a test bed for the different technologies being developed for the aperture arrays that make the front-end of the SKA1-LOW instrument. These include the antennas, LNAs, RFoF electronics and digital processing, but also the EM simulation software being developed for the project [16]. The AAVS0 is located at the MRAO at Lords Bridge and has 16 antennas arranged in a random configuration (see Fig. 4). The choice of a random configuration, in order to mimic the configuration proposed for SKA, is based on a minimization of the detrimental effects of both side lobes [5] and mutual coupling [4, 6]. The exact configuration of AAVS0, which is identical to the test array installed in Western Australia, AAVS0.5 [17], has the same randomisation and density as the SKA1-LOW instrument. The average spacing between elements is half a wavelength at 77 MHz.

The AAVS0 array is shielded from the soil by a 15 m diameter circular metallic mesh with a wire thickness of 2.5 mm and a pitch of 2.5 cm. The simulation work has shown that the optimum for SKA1-LOW is actually a coarser mesh which will save cost while still improving the polarization response of the antennas [4, 18]. The array processing back-end is located in a metallic hut, located 15 m from the edge of the ground plane (see Fig. 5). The close proximity of the hut intends to limit the length of the copper cables (2 per antenna) carrying power and RF signal. Both the metallic hut, the copper cables and the soil had to be included in the EM modeling as discussed below in order to achieve good agreement between measurements and simulations at the low end of the frequency band.

An RF receiver chain was built for AAVS0 as shown in Fig. 6. After the antenna and LNA, the 32 dual polarization signals were fed to the metal bunker using 30 m long coaxial cables. These signals were then terminated in pre analogue to digital (PREADU) boards which served the purpose of filtering and amplifying the signals as well as allowing a switched auxiliary noise input. The high RF enviroment at Lord's Bridge meant that FM filters had to be used to block the signal range 88-108 MHz, otherwise the receiver system would saturate. The total gain including the LNAs, could be modified from ~55 dB to ~85 dB in 0.5 dB steps with the PREADU gain range being nominally 12 – 43 dB. The ouput of the PREADU was filtered by 400 MHz anti-aliasing filters. Outputs of antenna #1 and #5 have been connected to a GPS-triggered spectrum analyzer in span zero mode [8], in order to measure the embedded-element patterns shown in section 6.B. As far as the array patterns in section 6.C are concerned, the output from all antennas was instead combined using an analog power combiner and then fed into the spectrum analyzer.

In the past, multiple tests were performed on the AAVS0 array, including coupling tests [19], polarization tests with an artificial source [20], and near field pattern test [21]. Astronomical sources have been instead used in [17] to charaterize the co-polar array beam of AAVS0.5. However, only with the help of the UAV system described in section 5, which provides higher dynamic range and measurement flexibility, we have been capable of performing accurate measurements of both embedded element patterns and array beam (co and cross-polar) in the whole operative frequency range.

4. EM MODELING OF AAVS0

The accurate EM modeling of the station beams of SKA1-LOW is crucial for the design and calibration of the system. Mutual coupling has a strong effect on the noise matching, beam shape, calibrability and ultimately the dynamic range of the instrument. Therefore a great deal of work has gone into building a EM simulation environment capable of handling the peculiarities of SKA, mainly the large station size (tens of wavelengths across) and the non-regularity of the antenna layout.

An in-house Method of Moments code is being developed for this purpose based on [16]. This code based on the Method of Moments and the Macro Basis Functions technique, will produce station beams in a matter of seconds that can be described with few coefficients to facilitate its fitting using a reduced number of available strong astronomical sources [22]. This code has been validated using available commercial software packages. The size of a SKA1-LOW station represents a large computational burden for the commercial packages, which are currently not practical for EM characterization of arrays much larger than AAVS0. We can however use the generic commercial codes to both validate our own codes and small test arrays. In this paper we have used one of these commercial software packages (FEKO [23]). The antenna model in FEKO was created by meshing a flat version of the antenna elements made of Perfect Electric Conductor surfaces to save simulation resources for the complete array. The same model simplification has been performed on the meshed ground plane (see Fig. 5)

The soil beneath the ground plane is also included in the model [4]. It is "visible" to the antenna elements, especially to those near the edge of the ground plane. We also included the presence of the feeding cables and the metallic hut (see Fig. 5). The hut is a trapezoidal prism which dimensions for the base (and top) are 2.78 x 1.83 $m^2$, the large wall is of 2.78 x 2.3 $m^2$, the small wall is of 2.78 x 2.06 $m^2$ and two more small walls covering the sides.

The simulations were performed on a cluster of four computers with 384 GB of RAM. The phase shifting caused by the LNAs and phase delays due to the cable length have been taken into account in FEKO. This has been done by adding the measured phase delay to each embedded element pattern in the script written in EDITFEKO to calibrate the total array response for each polarization (more information on this in section 6).

## 5. UAV-BASED MEASUREMENT SYSTEM AND FLIGHT STRATEGIES

The UAV system [11] is an artificial test-source based on a commercial hexacopter, which has been equipped with RF devices such as a synthesizer, a balun and a dipole antenna (length is selected according to the specific test frequency). After the take off, the UAV follows an autonomous GPS-guided navigation according to a pre-programmed flight path. The system allows for both co- and cross-polarization radiation pattern measurements along specific cuts. During the AAVS0 campaign, this task has been accomplished performing constant-height flights at 100 m and 150 m for the operative frequencies 50-250 MHz and 350 MHz, respectively, in order to satisfy the far-field condition. In particular, the UAV bearing/yaw angle has been sequentially set to either 0° or 90° from North in order to excite the two array polarizations alternatively. It should be noted that only four pattern cuts covering the full 90 Deg field-of-view can be performed within the limited flight duration of about 10 minutes, considering a UAV speed of about 3 m/s. Additional polarization studies have also been carried out performing slow rotations of the UAV around its vertical axis (spin flights). Moreover, 2D maps of the array beam have been recorded performing specific flights along a cartesian raster. A total amount of approximately 30 flights has been carried out during the whole campaign.

During the flight, a continuous-wave signal is transmitted in the frequency range between 50 and 350 MHz. The relevant signal at the output of the AAVS0 system (see section 3) is recorded together with a time-stamp provided by a GPS receiver. The UAV orientation and its real position in the sky are accurately measured by means of the on-board Inertial Measurement Unit and an additional differential GNSS receiver (centimetre accuracy), respectively. All these data are inserted as inputs of the post-processing scheme [14] to obtain the radiation patterns shown in section 6.

A side-application of the UAV system is the photogrammetry survey to determine both the element positions and orientations. This can be a very useful tool during the deployment phase of a radio telescope such as SKA1-LOW which can provide a snapshot of the array element positions at time 0. Through the photogrammetry, it is possible to get information about the 3D coordinates of an object and also to generate 3D textured models, Digital Surface Model (DSM) and orthophotos starting from images. During the AAVS0 campaign, two flights with a digital camera (Sony Nex 5) mounted on the UAV were performed above the array at two different heights (10 m and 20 m, see Fig. 3). The flights were planned to have a perfect photogrammetric coverage of the area with 60-80% overlay. The acquired images were processed with a commercial software (PhotoScan) that can perform images alignment to reconstruct the 3D point cloud of the area. To georeference the images, 14 markers were placed around the array (Fig. 7) and then their position were surveyed through a topographic approach using a total station and a prism. The 3D point cloud was then used to create the mesh and extract the DSM and the orthophoto of the area. These products were finally used to compute the 3D coordinates of the antenna centers with a high accuracy (~$\sigma_{xy}$ = 0.1 cm and $\sigma_z$ = 1.5 cm). The measured antenna positions were used as inputs of the EM model described in section 4.

## 6. RESULTS AND DISCUSSION

The UAV system opens a plethora of possible scenarios to be analyzed. In this section we divide the configurations we tested in three sub-sections: A) single-element, when the antenna is isolated with respect to other antennas; B) embedded-element, when the antenna receives the signal with all the other antennas present and loaded with the LNAs; C) array, when the signals received by all the antennas are summed to form the array pattern. For both the single- and embedded-element cases, the maximum measured value has been aligned to the simulated one in order to allow for a direct comparison. On the other hand, the array patterns have been normalized to the maximum. The amplitude calibration of the measurement system reported in [13] was not performed during the AAVS0 campaign.

It should be mentioned that the overall UAV-based system accuracy has not yet been verified against other measurement data because, as mentioned in the introduction, it is cumbersome if not impossible to reproduce the same measurement conditions (i.e. antennas on the ground radiating at VHF frequencies) in anechoic chambers or with other outdoor test ranges. An accuracy budget has been estimated in [13] by analyzing all the error contributions due to position, orientation and RF part uncertainties. The resulting value is in the order of 0.5 to 1 dB (between measurement and simulations) for the co-polar component, which is consistent with the discrepancy values observed in the previous validation campaigns performed on simple biconical and log-periodic antennas [11]. Similar discrepancy levels have been generally obtained in this work. More specific comments are reported in the corresponding sub-sections.

*A. Single-element measurements*

The single dual-polarized SKALA element was placed about 60 m from the spectrum analyzers, which have been used to directly record the data from the output of the LNAs. It should be noted that both antenna ports were simultaneously recorded in order to increase the overall measurement efficiency. The antenna was located on top of a 2.4 by 2.4 m mesh ground plane. The pattern measurements have been performed in the principal planes $\varphi=0°$ (East-West direction, see Fig. 7) and $\varphi=90°$ (North-South direction).

The obtained measured and simulated results at 350 MHz are reported in Figs. 8 for the branch of the SKALA oriented to the North-South direction. Therefore, the $\varphi=0°$ and $\varphi=90°$ cuts corresponds to the H- and E-planes, respectively. The co-polar (cross-polar) patterns have been obtained setting a North-South (East-West) orientation of the UAV-mounted test source dipole, corresponding to a bearing angle, i.e. the orientation of the source dipole with respect to the North-South direction, of 0° (90°).

As far as the co-polar measurements are concerned, the discrepancy between measurement (solid line) and simulation (dashed line) is generally within 1 dB for the E plane and 0.5 dB for the H plane. The cross-polar measurements show a higher discrepancy with respect to the corresponding simulations due to a combination between the lower Signal-to Noise ratio and the geometrical uncertainties of both the AUT and the test source with particular reference to *i)* the AUT angular position accuracy on the measurement field, *ii)* the alignment of the source dipole on the UAV, and *iii)* limited accuracy of the UAV orientation measured by the on-board Inertial Measurement Unit, which is in the order of 2°. Nevertheless, it should be noted that the overall cross-polar pattern shape is also very consistent between measurement and simulations. Similar results have been obtained for the orthogonal antenna.

The measured radiation patterns at 50 MHz are compared to the simulated ones in Fig. 9. The overall agreement is very good. Both the same scanning strategy and reference system described above have been adopted. However, at 50 MHz, due to the limited size of the largest dipole of SKALA, part of the current is confined in the boom of the antenna and part in the bottom dipole, which causes a partial rotation of the polarization axis. The measured cross-polarization value, which is higher than the co-polar one, suggests that such a rotation is larger than 45°.

A specific scan strategy has been adopted in order to further characterize the SKALA polarization behavior at 50 MHz. The UAV-based test source, positioned at zenith, performed a slow rotation around its vertical axis. Figure 10 shows the antenna gain at zenith for both antenna polarizations as a function of the UAV bearing angle. It can be observed that the maximum response for the NS alignment (black curves) is located at about 60° from the co-polar axis defined above (bearing 0°). The measured data are consistent with the simulations. The discrepancy on the angular position of the nulls can be also attributed to the geometrical uncertainties discussed above. Figure 10 also shows that the behavior of the antenna oriented along EW (gray curves) is shifted of about 90°. This means that, despite the principal polarization of the two antenna branches are not aligned with respect to the adopted coordinate system, they are electromagnetically orthogonal to each other. In more detail, the level of orthogonality between the two polarizations can be expressed in term of the Intrinsic Cross Polarization Ratio (IXR) [24]. According to the procedure reported in [25], the IXR level at zenith can be directly estimated from the data in Fig. 10. In particular, the relative angular distance between two adjacent minima, which is about 86° (instead of 90°), produces an estimated IXR level of about 30 dB. It should be noted that such a measured level is also affected by the UAV bearing angle accuracy discussed above. The same measurement strategy has been also adopted at the higher frequencies showing similar IXR values.

*B. Embedded-element measurements*

Embedded-element patterns of a few elements of the AAVS0 demonstrator have been measured and compared to simulations in order to quantify the confidence level of the SKALA EM models in the random array configuration. It should be emphasized that this aspect is crucial in the context of a model-based calibration of the telescope.

As described in section 3, the spectrum analyzers have been connected after the PREADU. Similarly to the single-element measurement described in section 6.A, two cuts at $\varphi=0°$ and $\varphi=90°$ have been considered for both co-polar and cross-polar orientation of the test source. The results at 350 MHz are reported in Fig. 11 and 12 for element #1 and #5, respectively. Only one antenna branch (oriented along NS) is shown. With reference to Fig. 4, element #5 is located in the center whereas element #1 is at the border. As expected, the embedded-element patterns show a slightly distorted behavior with additional ripple when compared to the single-element ones reported in Fig. 8. It should be observed that such phenomena are clearly visible in both measurements and simulations, which are in very good agreement. For both co-polar and cross-polar, the observed discrepancy levels are quite similar to the ones obtained for the single-element case. This result suggests that the developed EM model (embedded-elements) can be used to predict the polarization characteristics of the overall array beam versus scan angle with an acceptable level of confidence.

In particular, the discrepancy between measurement and simulation is slightly higher for element #1. Such an element is located at the border of the AAVS0 configuration. Hence, it is more sensitive to both ground parameter (metallic mesh deformations and soil permittivity) and the overall adjacent environment e.g. vegetation and other infrastructures.

Similar considerations can be applied to the embedded-element patterns at 50 MHz, which are reported in Fig. 13 and 14 for element #1 and #5, respectively. The cross-polar data are again higher that the co-polar ones of about 2 dB for the same reasons discussed in section 6.A. Therefore, it can be concluded that the polarization characteristic of the SKALA element is maintained also within the array environment. At this low frequency, the slightly higher noise level on the data is related to the increased mismatch loss at both the test source and SKALA ports. Finally, it should be observed that the SKALA element exhibits a lower directivity at 50 MHz (i.e. a wider beam). Therefore, at this lower frequency, its behavior in the array is definitely more affected by the non-idealities of the surrounding environment (adjacent elements, cables, ground mesh, soil), This aspect is in turn responsible of the slightly higher discrepancy between measurements and simulations (generally within 2-3 dB).

*C. Array measurements*

As introduced in section 3, the 16 outputs of the PREADU (branch NS of each element) have been summed to obtain the AAVS0 radiation pattern at zenith using an analog power combiner from Mini Circuits [26]. Such a test solution is definitely simpler to implement with respect to the digital correlator described in [14]. Nevertheless, it allows for a verification of the full array beam, although in a single specific scan condition.

The obtained array patterns at 50 MHz, 150 MHz, 250 MHz and 350 MHz are reported in Fig. 15. Both the beamwidth variation versus frequency and the angular position of the side lobes are very consistent between measurement and simulations. A slight beam tilt toward the positive zenith angles as well as an increased secondary lobe level in the negative angular region can be observed at all frequencies. These discrepancies could be due to a possible inclination of the array concrete base that has not been characterized during the campaign. Moreover, the surrounding environment could have produced very small spurious reflections.

It should be pointed out that the simulated data in Fig. 15 have been computed taking into account the real phase shift of each channel of the RF front-end. In particular, the transmission coefficient of every cable and channel was measured and accounted for in the simulation i.e. 16 (30 m) coaxial cables from the LNAs to the receiver box in the hut, 16 receiver channels, 16 (1 m) short coaxial cables connected from the output of the receiver box to the input of the power combiner, and the power combiner itself. In other words, the overall analog beam forming network has been measured and taken into account in the simulation, Therefore, the present array beam measurement can be certainly considered as an aggregate (even if not complete) verification of all the embedded element patterns.

The "uncalibrated simulation" at 350 MHz i.e. with equal phase shift on all channels is also reported in Fig. 15 with dotted line in order to show the effect that phase errors in the combining network produce on the array radiation pattern. The higher frequency was selected for this study owing to its higher sensitivity to phase errors. The benefits of the above mentioned phase correction are evident. In particular, they are more pronounced in the region containing the first secondary lobes, whereas the main beam and far side lobes are less affected.

Through the side lobes of the SKA1-LOW stations, undesired power from sources located away from the main beam will be captured contributing to raising the noise floor of the instrument. The effect of these side lobes in astronomical interferometric observations, such as the ones that SKA1-LOW will routinely do, can therefore be a limiting factor for the dynamic range of the instrument. In consequence, it is necessary to have a very good understanding of the spectral, spatial location and stability of these secondary lobes both in the design and calibration phase.

As described in [27], it can be shown that a random array exhibits two clearly different regions, a first "coherent" region near the main beam of size in number of side-lobes defined as

$$P = \sqrt{\frac{N}{\pi}} \tag{1}$$

(based on a Nyquist criterion of aperture sampling); and a second one beyond that point where the contribution of different antennas add up on average in power rather than in amplitude, which is called "non-coherent". In [27], it is also noted that the actual size of the "coherent" region may be actually smaller than the value predicted in (1) due to limitations of the average density in this type of arrays. For the AAVS0 configuration, P is equal to 2.2. The corresponding coherent aperture-like region where one may expect a stronger influence of phase deviations on the pattern is marked with the gray shaded rectangles in Fig. 15. The observed sensitive regions of the array pattern with respect to phase error in the combining network are clearly consistent with the theoretical prediction.

Finally, a raster measurement of the AAVS0 array pattern at 350 MHz has been performed to obtain the 2D plots in Fig. 16. A sequence of evenly-spaced linear scans has been performed covering a flat area in the array far-field, with constant test-source orientation (North-South). The size of the raster has been limited to the first secondary lobes owing to the available flight time duration. The measured data have been interpolated (griddata) in order to obtain the array power pattern represented in the uv plane with satisfactory agreement between measurement and simulations. The difference between the two plots in Fig. 16 has been reported in Fig. 17, according to the logarithmic-difference error definition in [28]. As expected, the obtained 2-D error levels are higher in the null regions of the array pattern. The change of error sign in the main beam region (light cyan and yellow areas) is consistent with the small beam tilt shown in Fig. 15. Such an effect is only present along the plane φ=0°.

## 7. CONCLUSIONS

This paper describes the beam modeling and measurement of a 16-element random array of ultra wideband Log-Periodic antennas (SKALA) dedicated to the development of technology and techniques for the SKA1-LOW instrument. We have shown good agreement between the electromagnetic simulations and measurements using a micro UAV system validating the antenna and array designs. Furthermore, we have described the measurement of a rectangular 2D patch in the far field of the array useful for wide field of view validation of modern all-sky radio arrays. Through the measurement of cuts of the array pattern we have verified the effect of un-calibrated phase on the inner side lobes of random arrays, as predicted by theory. The final discrepancy reached in the AAVS0 antenna patterns between the UAV measurements and the numerical results is still lower than the target values of the SKA. However, these tests proved that the UAV-base technique allows an accurate knowledge of the antenna/array patterns. These patterns could then be used in a larger calibration pipeline for SKA. Future activities will be addressed to combine the strengths of the UAV with other complementary methods to calibrate the telescope.

The next campaigns and developments will focus on an array with a digital back-end to exploit the possibility of measurements of the complex field pattern as well as near field patterns. Furthermore, we envisage the use of a similar technique with the AAVS1 system, a 400-antenna element prototype array currently under development in Western Australia.

## 8. ACKNOWLEDGMENTS

The authors would like to acknowledge their SKA colleagues for numerous discussions and constructive input about the topic. This research was supported by the Science & Technology Facilities Council (UK) grant: *SKA, ST/M001393/1* and the University of Cambridge, UK. This work is also part of the activities included in the "Advanced calibration techniques for next generation low-frequency radio astronomical arrays" project supported by the 2014 TECNO INAF funds.

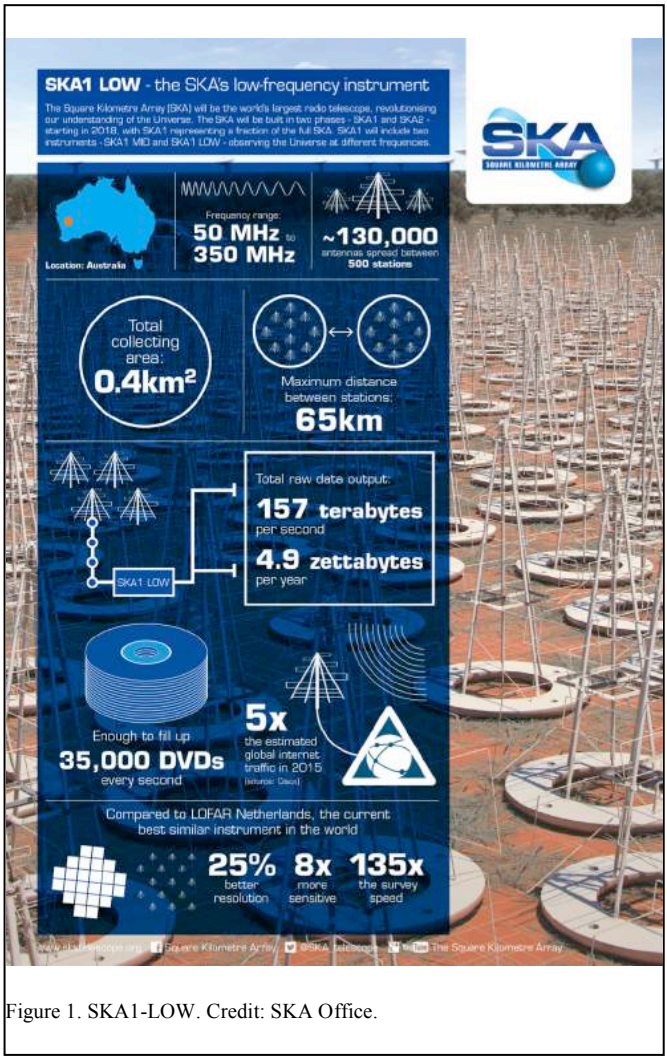

Figure 1. SKA1-LOW. Credit: SKA Office.

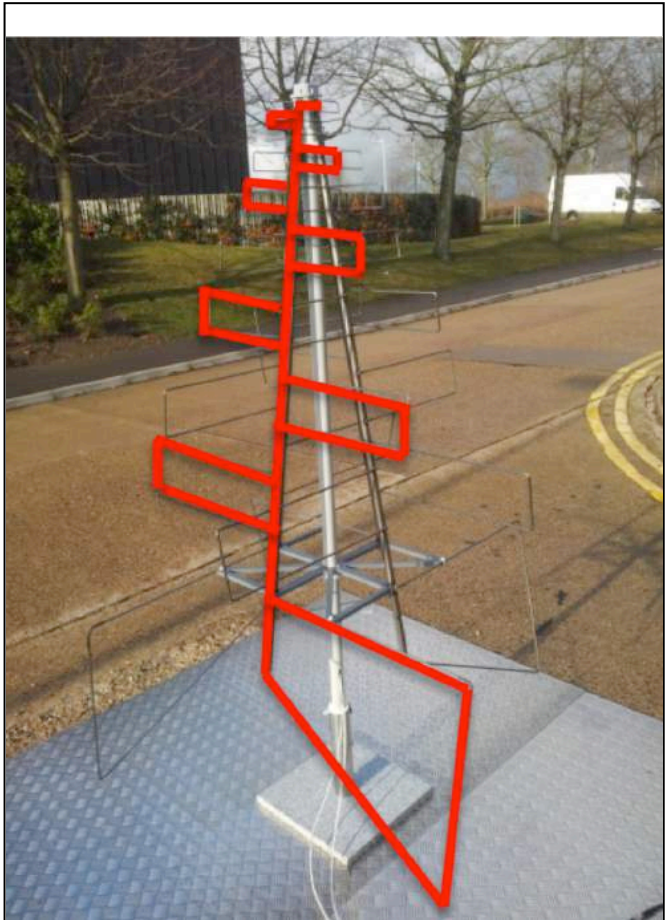

Figure 2. SKALA-1. 1 arm highlighted in red.

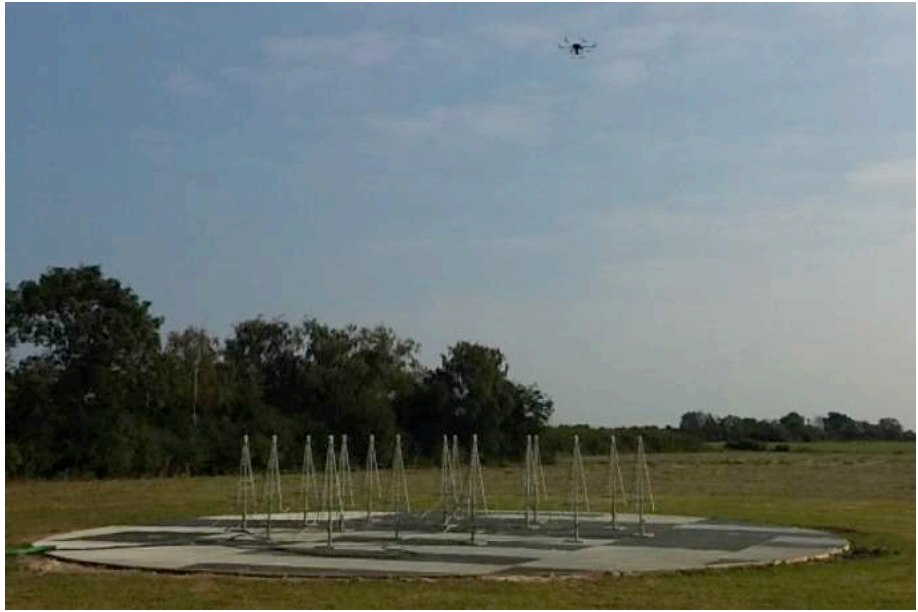

Figure 3. UAV flying over the AAVS0 array, deployed at the Mullard Radio Astronomy Observatory, Lords Bridge, Cambridge.

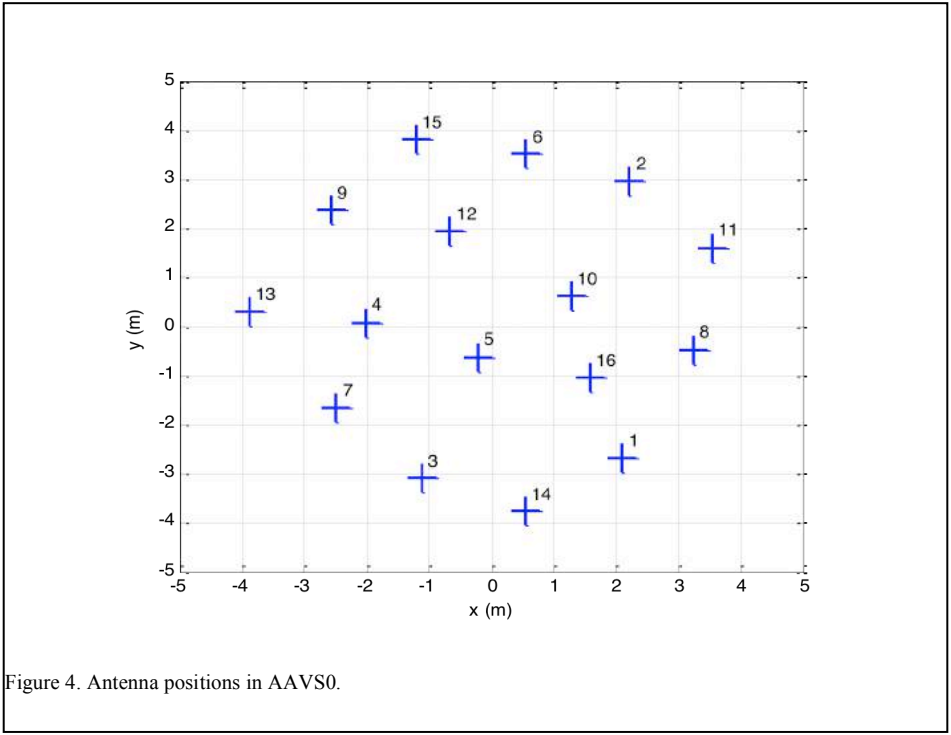

Figure 4. Antenna positions in AAVS0.

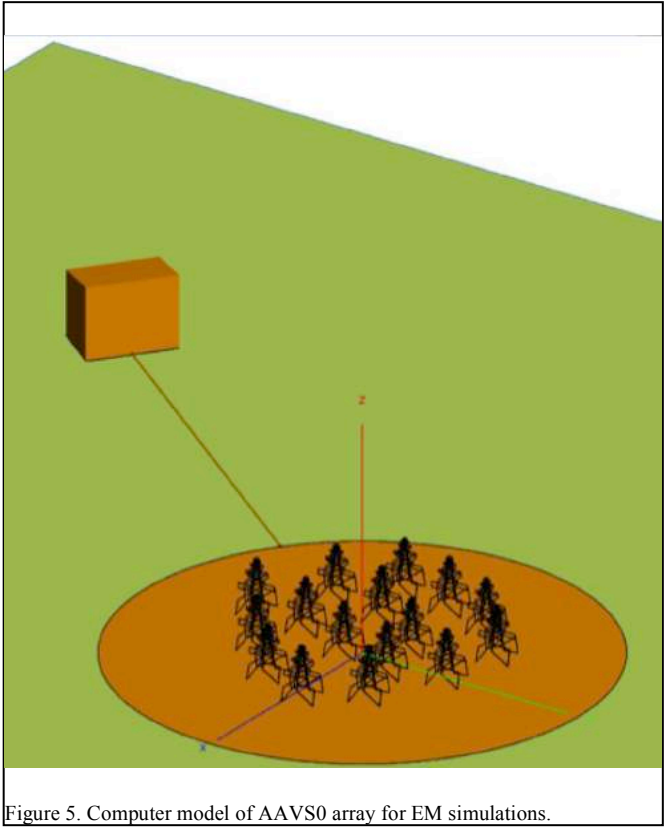

Figure 5. Computer model of AAVS0 array for EM simulations.

Figure 6. AAVS0 RF system.

Figure 7. The AAVS0 element positions (red markers) were measured by the UAV photogrammetry system with ~1cm accuracy. The blue markers

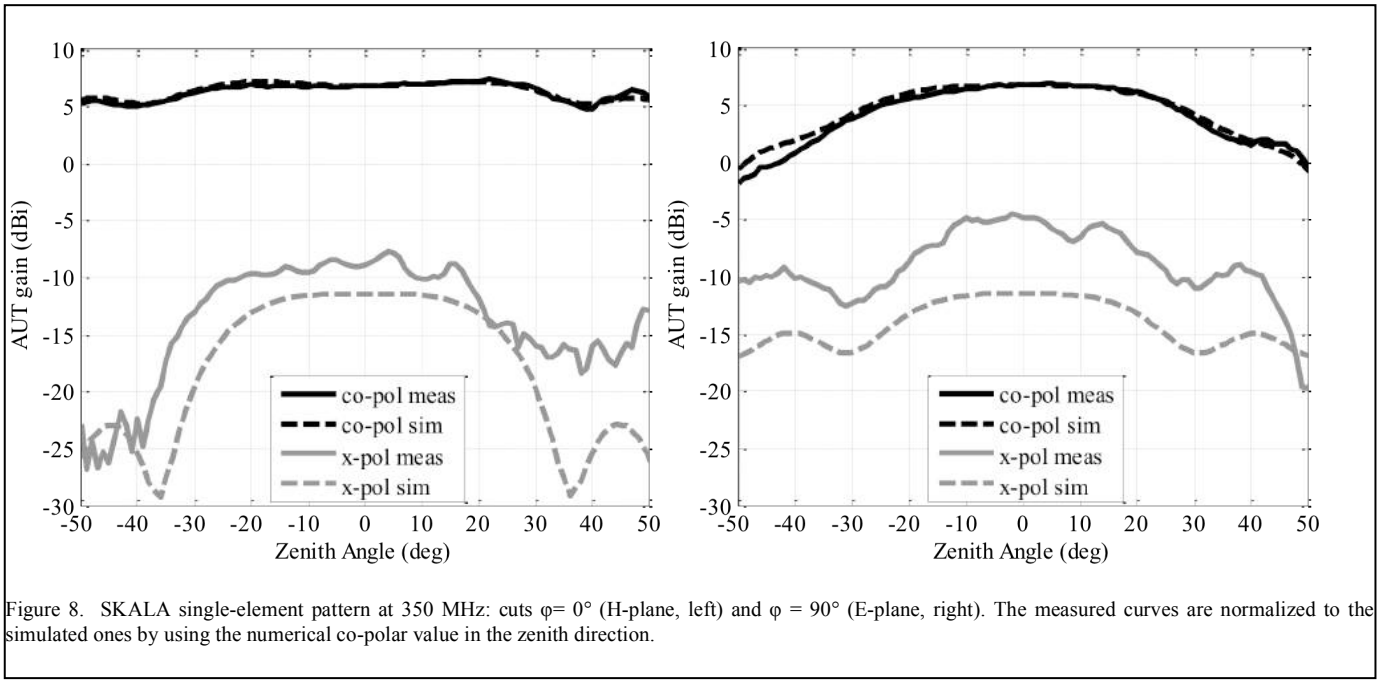

Figure 8. SKALA single-element pattern at 350 MHz: cuts φ = 0° (H-plane, left) and φ = 90° (E-plane, right). The measured curves are normalized to the simulated ones by using the numerical co-polar value in the zenith direction.

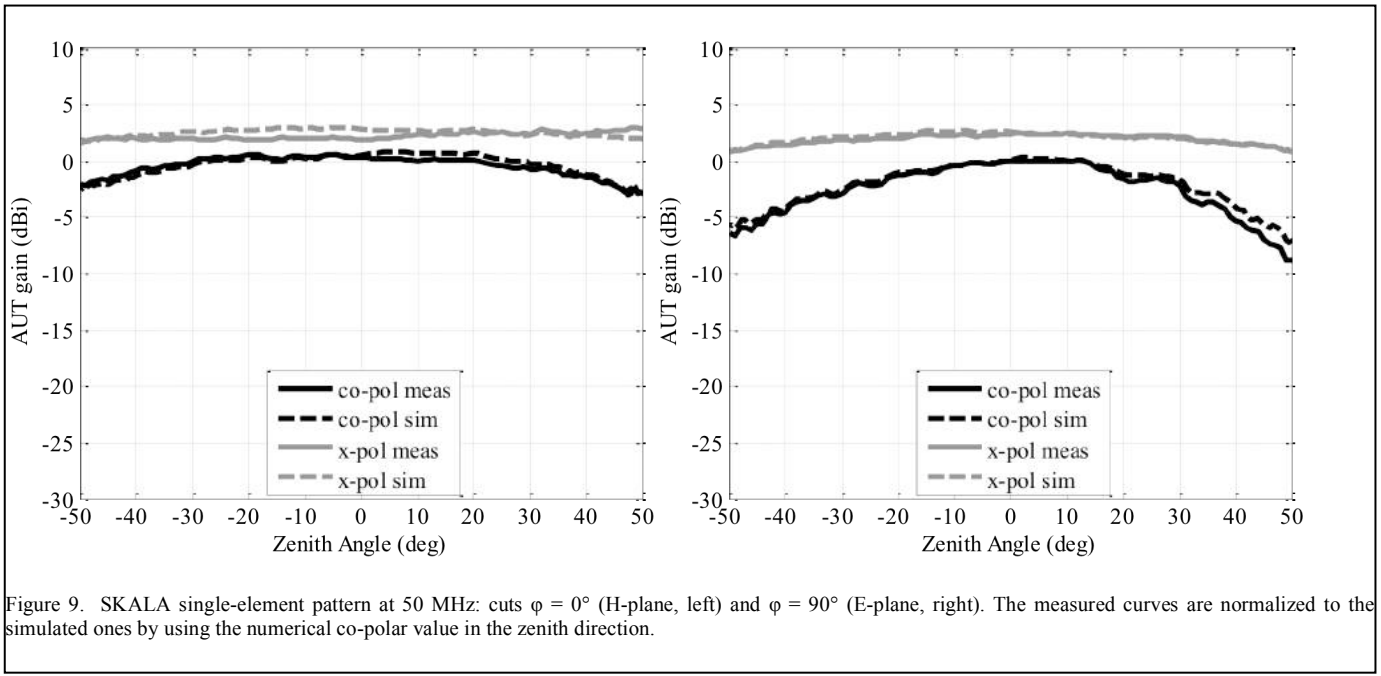

Figure 9. SKALA single-element pattern at 50 MHz: cuts φ = 0° (H-plane, left) and φ = 90° (E-plane, right). The measured curves are normalized to the simulated ones by using the numerical co-polar value in the zenith direction.

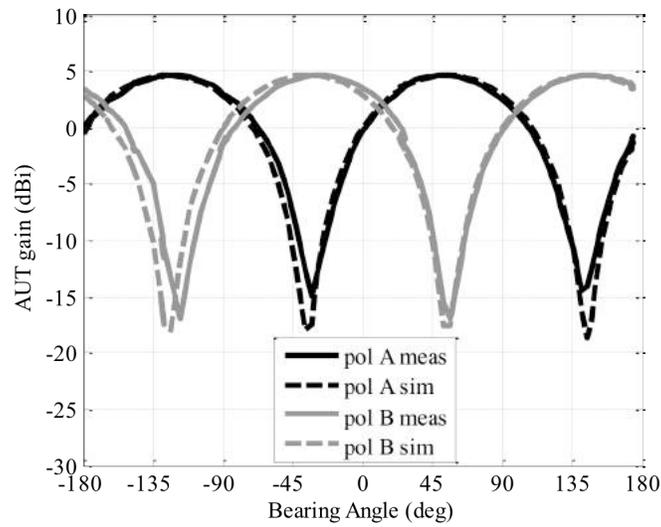

Figure 10. SKALA single-element pattern at 50 MHz: gain at zenith as a function of the UAV bearing angle. The measured curves are normalized to the simulated ones by using the numerical co-polar value in the zenith direction.

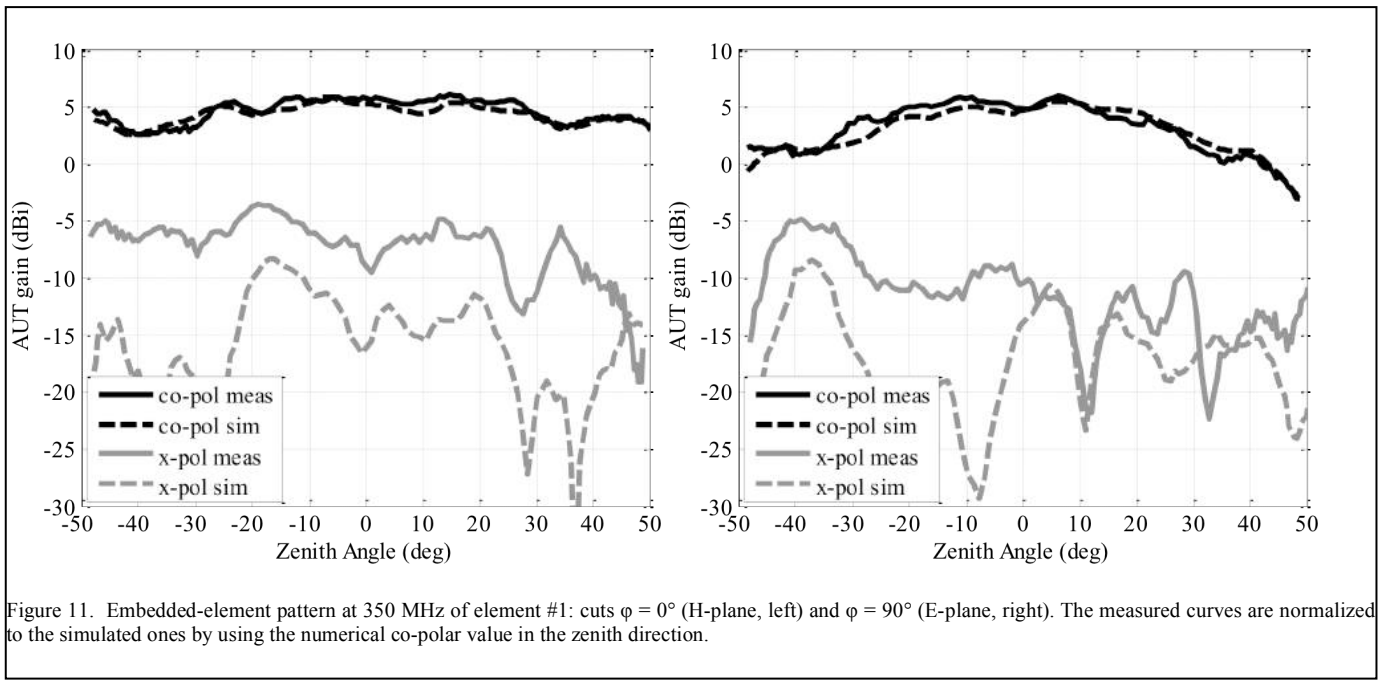

Figure 11. Embedded-element pattern at 350 MHz of element #1: cuts φ = 0° (H-plane, left) and φ = 90° (E-plane, right). The measured curves are normalized to the simulated ones by using the numerical co-polar value in the zenith direction.

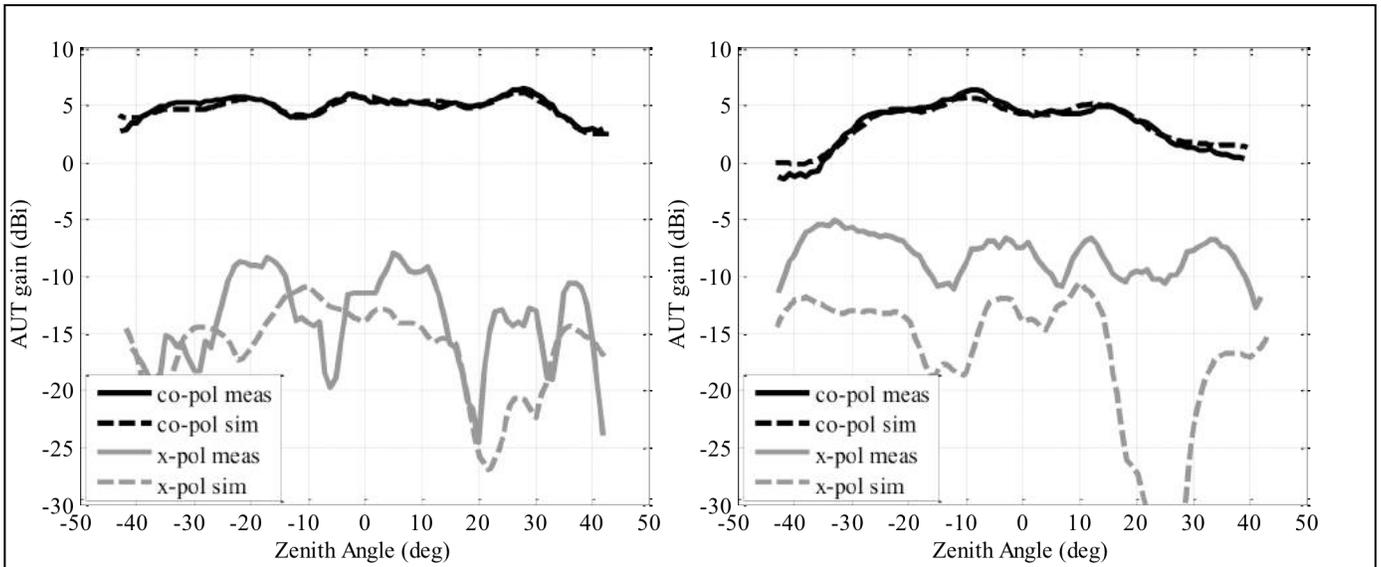

Figure 12. Embedded-element pattern at 350 MHz of element #5: cuts φ = 0° (H-plane, left) and φ = 90° (E-plane, right). The measured curves are normalized to the simulated ones by using the numerical co-polar value in the zenith direction.

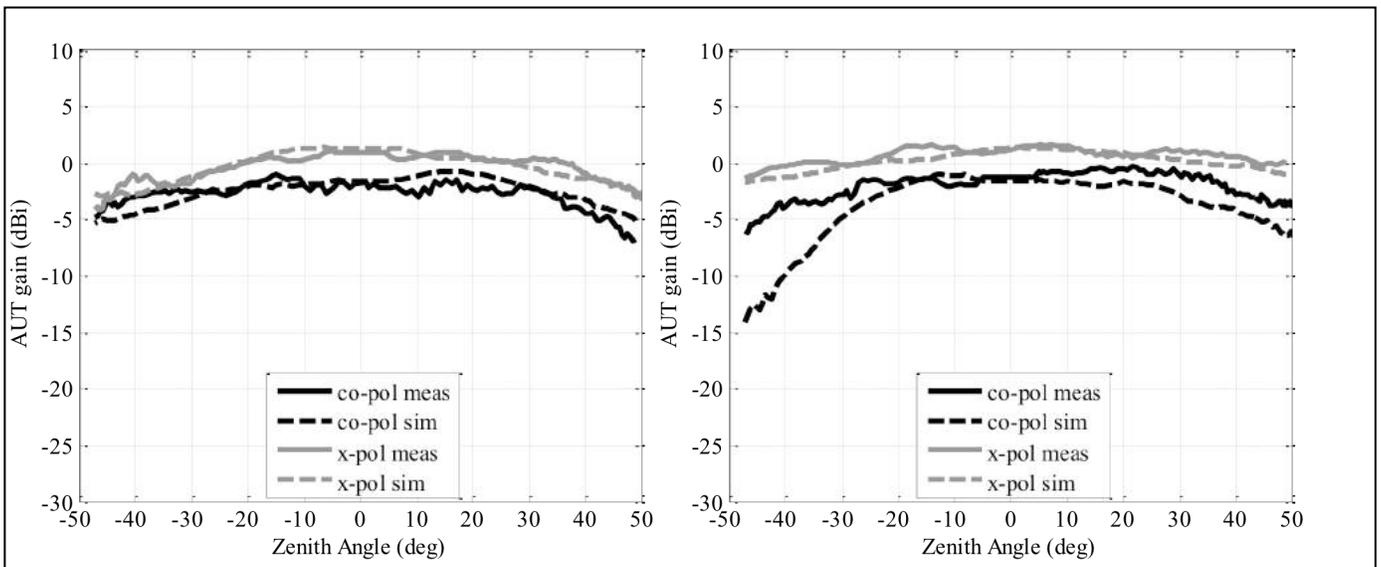

Figure 13. Embedded-element pattern at 50 MHz of element #1: cuts φ = 0° (H-plane, left) and φ = 90° (E-plane, right). The measured curves are normalized to the simulated ones by using the numerical co-polar value in the zenith direction.

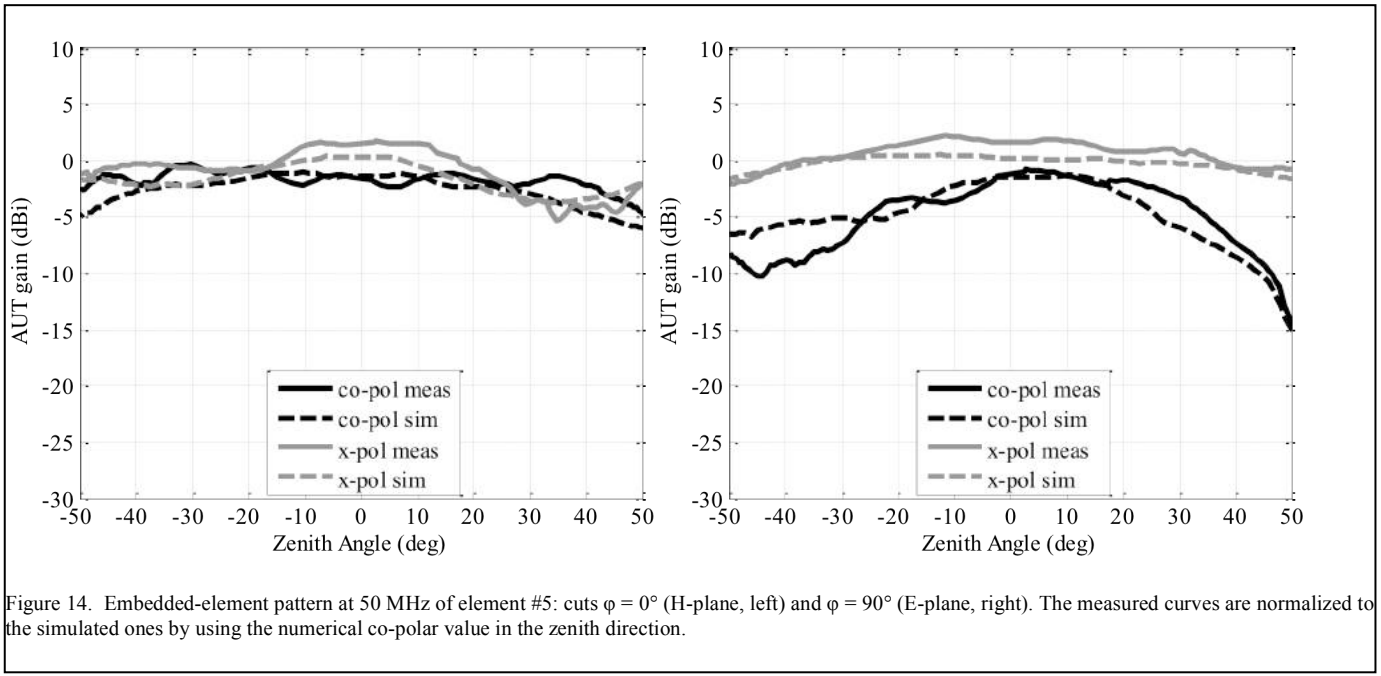

Figure 14. Embedded-element pattern at 50 MHz of element #5: cuts φ = 0° (H-plane, left) and φ = 90° (E-plane, right). The measured curves are normalized to the simulated ones by using the numerical co-polar value in the zenith direction.

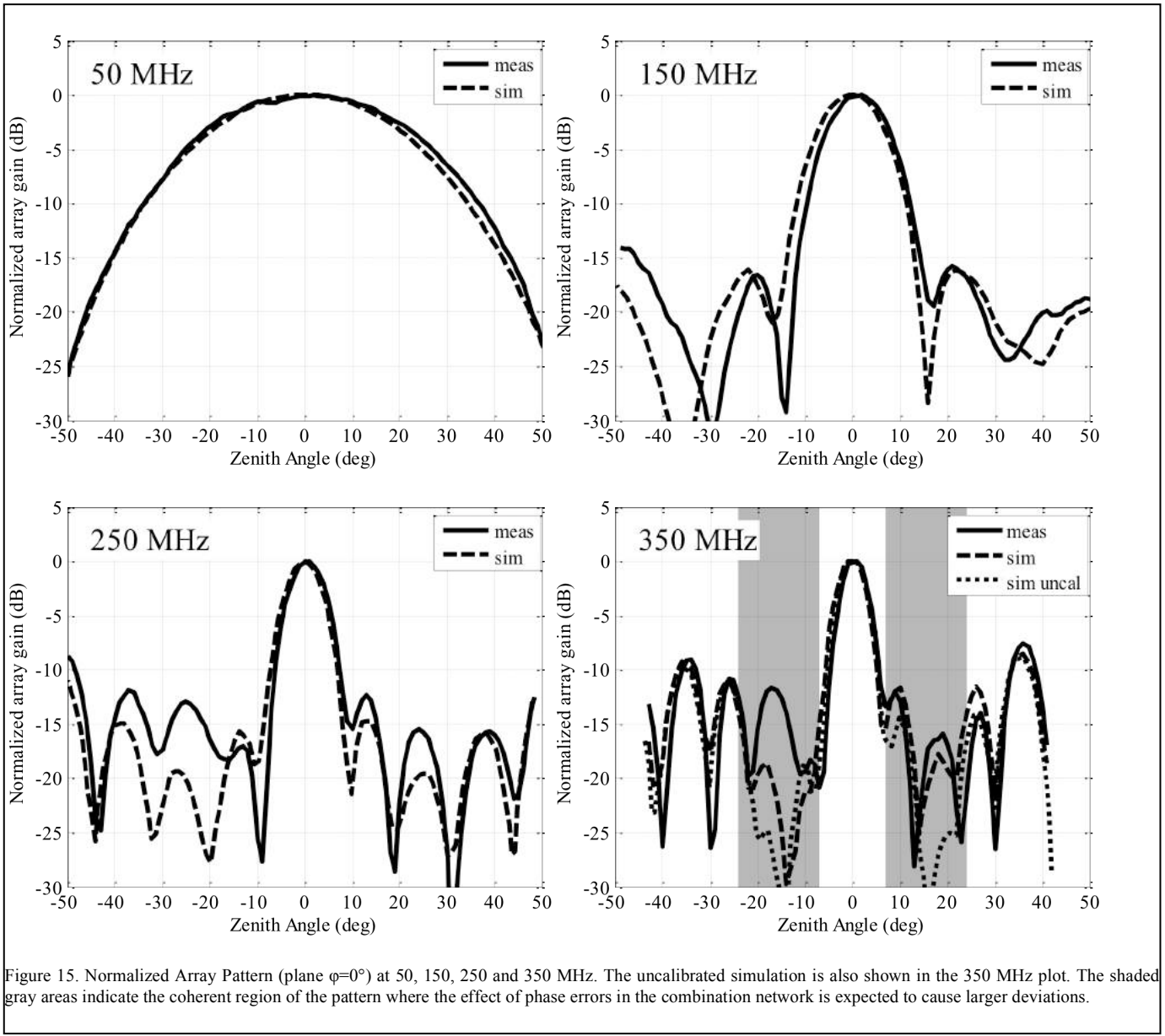

Figure 15. Normalized Array Pattern (plane φ=0°) at 50, 150, 250 and 350 MHz. The uncalibrated simulation is also shown in the 350 MHz plot. The shaded gray areas indicate the coherent region of the pattern where the effect of phase errors in the combination network is expected to cause larger deviations.

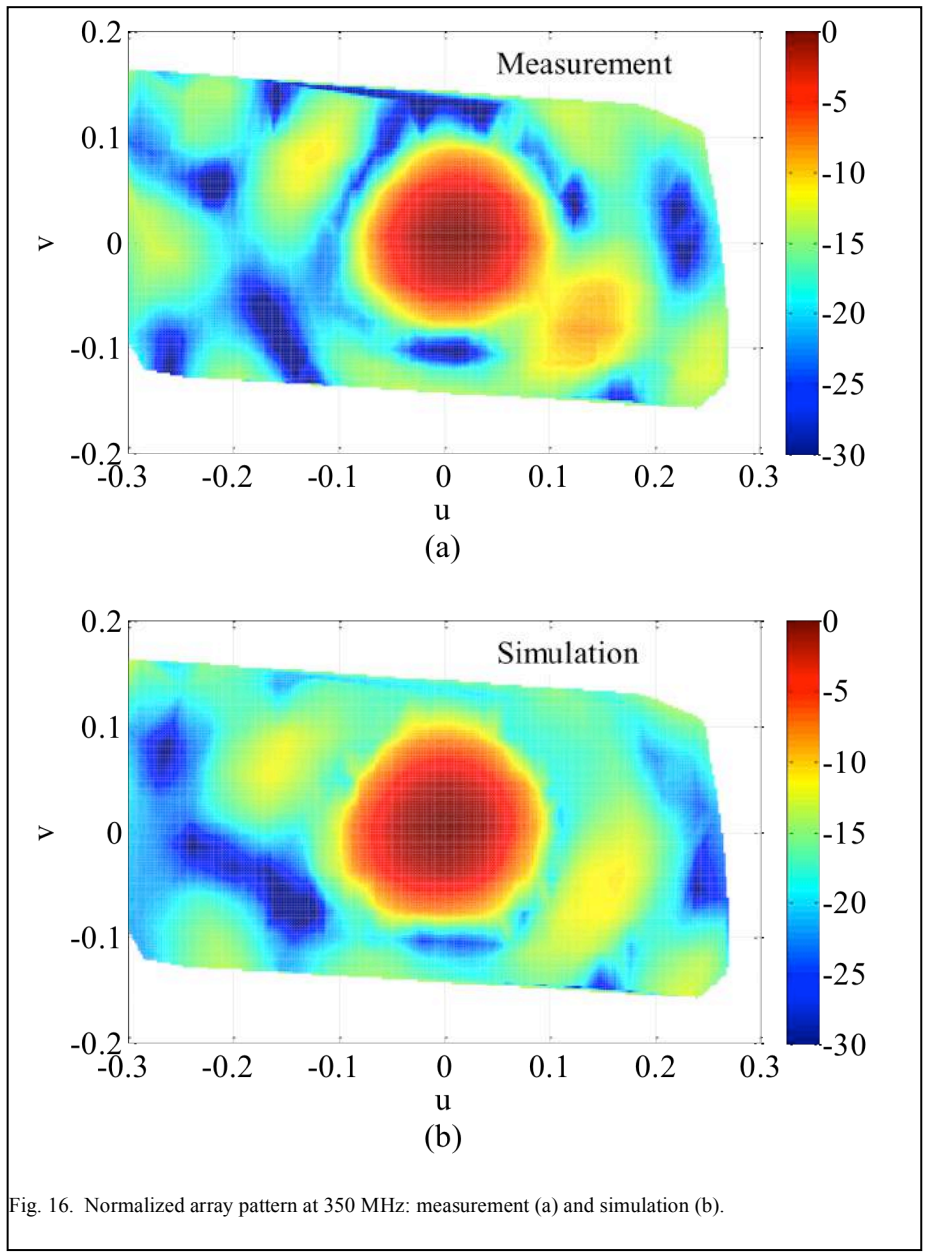

Fig. 16. Normalized array pattern at 350 MHz: measurement (a) and simulation (b).

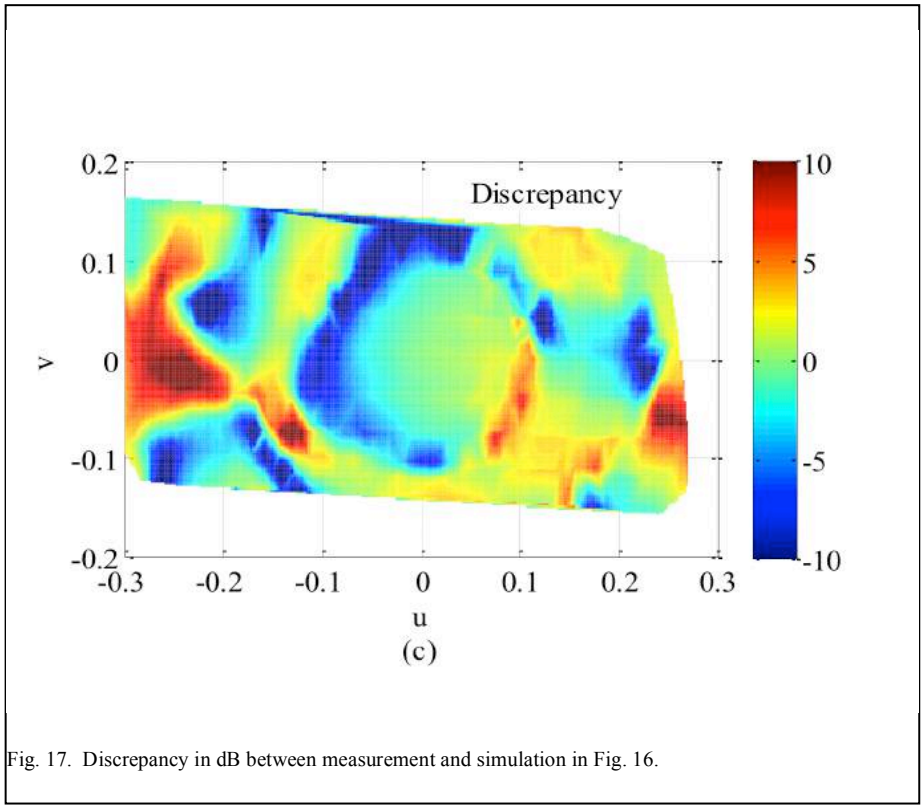

Fig. 17. Discrepancy in dB between measurement and simulation in Fig. 16.